# Spin-Charge-Lattice Coupling through Resonant Multi-Magnon Excitations in Multiferroic BiFeO$_3$

M. O. Ramirez,[1] A. Kumar,[1] S. A. Denev,[1] Y. H. Chu,[2,3] J. Seidel,[2,3] L. Martin,[2] S-Y. Yang,[2] R. C. Rai,[4,] X. Xue,[4] J. F. Ihlefeld,[1,2] N. Podraza,[1] E. Saiz,[2] S. Lee,[5] J. Klug,[6,7] S. W. Cheong,[5] M.J. Bedzyk,[6,7] O. Auciello,[7] D. G. Schlom,[1] J. Orenstein,[3] R. Ramesh,[2,3], J. L. Musfeldt,[4] A. P. Litvinchuk[8] and V. Gopalan[1]

[1]*Dept. of Materials Science and Engineering and Materials Research Institute, Pennsylvania State University, University Park, PA 16802 USA*

[2] *Materials Sciences Division, Lawrence Berkeley National Laboratory, Berkeley, California 94720 USA*

[3] *Dept. of Physics, University of California, Berkeley, California 94720-1760, USA*

[4] *Dept. of Chemistry, University of Tennessee, Knoxville, TN 37996 USA*

[5] *Rutgers Center of Emergent Materials and Dept. of Physics and Astronomy Rutgers, The State University of New Jersey, 136 Frelinghuysen Road, Piscataway, NJ 08854-8019 USA.*

[6] *Materials Research Center, Northwestern University, Evanston, IL 60208*

[7]*Materials Science Division, Argonne National Laboratory, Argonne, IL 60439*

[8] *Texas Center for Superconductivity and Advanced Materials and Department of Physics, University of Houston, Houston, TX 77204-5002, USA*

Spin-charge-lattice coupling mediated by multi-magnon processes is demonstrated in multiferroic BiFeO$_3$. Experimental evidence of two and three magnons excitations as well as multimagnon coupling at electronic energy scales and high temperatures are reported. Temperature dependent Raman experiments show up to five resonant enhancements of the 2-magnon excitation below the Neel temperature. These are shown to be collective interactions between on-site Fe *d-d* electronic resonance, phonons and multimagnons.

Spin-charge-lattice coupling is central to the current interest in multiferroic materials. In particular, the coupling between magnetization and polarization can lead to cross-coupled phenomena such as magnetic control of polarization and electrical control of magnetism.[1-2] Multiferroic materials with both polar and magnetic order parameters usually show a relatively low-symmetry crystal structure due to the absence of both time and space inversion symmetries; hence a strong interaction between the low-lying magnetic and lattice excitations can occur, leading to rich new physics phenomena. One widely studied mechanism for spin-charge coupling is through electromagnons, which are single-magnon spin waves that are excited by an ac electric field.[3-6] A relatively unexplored mechanism in multiferroics is through multi-magnon excitations, which is the subject of study here. Multimagnons are known to be very sensitive to temperature effects and phase transitions.[7]

Bismuth ferrite, $BiFeO_3$, the focus of this study, has a robust ferroelectric polarization (~100 $\mu C/cm^2$) at room temperature,[8] that is the largest among known ferroelectrics. At room temperature (RT), $BiFeO_3$ is a rhombohedrally distorted ferroelectric perovskite with space group *R3c* and a Curie temperature, $T_C$ < 1100 K.[9] It also shows a *G*-type canted antiferromagnetic order below Neel temperature, $T_N$ < 640 K, and, in the bulk, an incommensurately space-modulated spin structure along $(110)_h$.[10] During the past few years, various physical properties of $BiFeO_3$ have been reported to show anomalies across $T_N$.[11-14] However, with the exception of neutron diffraction, none of these properties, such as optical birefringence, absorption, X-ray diffraction, and second harmonic generation have cleanly distinguished the magnetism from the polar contribution in this material. This makes the study of coupled phenomena challenging. Here we show that combining Raman scattering and linear optical spectroscopy of multi-magnons, phonons, and electronic energy levels can reveal spin-charge-lattice coupling in this system.

For our experiments, 4.5 µm $BiFeO_3$ film on (110) $DyScO_3$ and $BiFeO_3$ single crystal were grown according to references 12 and 14. The thick films were relaxed, and have $(001)_p$ pseudocube-on-pesudocube epitaxial geometry, with trigonal $C_{3v}$ crystal structure. Raman spectra were recorded under excitation at 488 nm in a back-scattering geometry by using a WITec alpha 300 S confocal Raman microscope equipped with a Linkam heating stage. For the crystallographic growth planes studied here, the optic axis is at an angle to the surface, and polarized Raman signal mixes multiple polarizability tensor components; hence Raman spectra were collected in unpolarized geometry. A Bruker Hyperion 3000 microscope coupled to a Bruker IFS 66/s FT-IR was employed for the mid-infrared absorption measurements. Optical transmittance measurements were performed using a Perkin Elmer Lambda-900 spectrometer between 4 and 730 K.

Though magnons energy are typically small (<100 cm$^{-1}$), Raman scattering for spin wave optical branches at energies higher than 500 cm$^{-1}$ have been previously reported in iron oxides due to their large unit cell, their corresponding multimagnon excitations located at energies larger than 1000 cm$^{-1}$.[15] Figures 1(a) and (b) depict the high frequency unpolarized Raman spectra for a 4.5 µm thick epitaxial film of BiFeO$_3$ collected at different temperatures. Further details on the temperature dependent Raman spectra under 1300 cm$^{-1}$ can be found in references 11, 14 and 16. Evidence for two- and three- magnon scattering at the shoulder of a previously reported two-phonon overtone can be observed, their energy values peaking at ~1530 and 2350 cm$^{-1}$, respectively. The two- and three- magnon excitations were identified by using the striking spectral similarity between BiFeO$_3$ and α-Fe$_2$O$_3$, the simplest case of an iron oxide containing only FeO$_6$ octahedra, where not only two-magnon scattering but also two-phonon overtones at very similar energies have been reported.[14-15] (Our films and crystals were confirmed to be phase-pure BiFeO$_3$ by using synchrotron X-ray diffraction, transmission electron microscopy, and scanning probe microscopy). In addition, to confirm these assignments, RT mid-infrared absorption experiments were performed in the same spectral range (Figures 1(c) and (d)). The two-magnon absorption band can be seen around 1550 cm$^{-1}$ (see Fig. 1(c)), though the strong contribution to the absorption spectra of the second order overtones in the 1400-1800 cm$^{-1}$ energy range precludes resolving this peak. At higher energies (Fig 1(d)), an absorption band centered at ~2415 cm$^{-1}$ is well resolved which reasonably matches the three magnons excitation from Raman scattering. These results not only support our previous assignments but also point out strong coupling interactions between the spin system (Fe-Fe exchange interaction) and electric dipole (ED) active excitations in BiFeO$_3$.

Based on the position of the two magnon scattering observed in the Raman spectra, an estimate of the nearest neighbor exchange J, described by the Heisenberg Hamiltonian, can be made. If spin deviations are created on adjacent sites in a perovskite antiferromagnet, the excitation frequency of the system arising from two magnon scattering is given by $J(2Sz-1)$ ) where $S$ refers to the individual spin on the magnetic site and $z$ represents the number of nearest neighbors to that site.[7] In Bismuth Ferrite, the spin on $Fe^{3+}$ site is $S = 5/2$ and the number of nearest neighbors around Fe cation is $z = 6$. The frequency shift of the two magnon peak to the first approximation is given by $\omega(2M) = J(2Sz-1) = 29J$. From experiments, the two magnon scattering peak is observed at 1530 cm$^{-1}$ for $BiFeO_3$ in both film and crystal . Thus, the value of exchange interaction is obtained as J ~ 1530/29 = 52.8 cm$^{-1}$ = 6.54 meV. This estimated value is in agreement with first principle calculations performed on $BiFeO_3$ using LDA + U method (For U = 5 eV,) where the |J| value is estimated to be 7.4 meV).[17]

As the samples were heated, a striking enhancement of the 2-magnon Raman bands was observed *only* within a ~10 K window around six different temperatures of $T_1^* = 365$ K, $T_2^* = 450$ K, $T_3^* = 490$ K, $T_4^* = 530$ K, $T_5^* = 600$ K and $T_N = 645$ K as seen from Fig. 1(a) and (e). Similar results were observed in single crystals. Unfortunately, the weak intensity and considerable linewidth of the two and three magnons excitations prevented a study of their temperature dependence in terms of frequency shift, broadening, and total intensity. Nonetheless, from the ratio between two-magnon (<1550 cm$^{-1}$) and two-phonon (<1260 cm$^{-1}$) excitations as a function of temperature, a clear plot of the temperature driven multimagnon enhancement can be obtained (Fig. 1 (e)). Since anomalies in the measured resistivity and induced magnetization were observed only at $T_N$, and not at any of $T_1^*$ to $T_5^*$, it rules out spin reorientation transitions except at $T_N$. Also, a lack of any anomalous lattice parameter changes at these temperatures,[18] suggests that these are not structural phase transitions. We note that the 2-phonon overtone - strongly enhanced due to the resonance with the intrinsic absorption edge, also couples strongly to $T_N$ as shown previously.[14] Hence, since the excitation energy ($E_{exc}$<2.54eV) is close to the bandedge, these anomalies appear to be (multi)magnon/phonon-assisted electronic resonances driven by temperature shifts. To further explore this possibility, we first probed the electronic structure of $BiFeO_3$ at high energies, *i.e* close to the band edge.

Temperature dependent linear absorption measurements (from T = 5 K to 730 K) and parameterization analysis were performed on a BiFeO$_3$ film on DyScO$_3$ (110). For the series of temperature dependent measurements, the dielectric function spectra ($\varepsilon_1$, $\varepsilon_2$) and corresponding complex index of refraction (N = n + ik) were extracted using a least squares regression analysis and a weighted error function, $\chi$, to fit the experimental ellipsometric spectra to an optical model consisting of a semi-infinite DyScO$_3$ substrate / 100 nm BiFeO$_3$ film / air ambient structure, where free parameters correspond to the parameterization of the BiFeO$_3$ dielectric function.[19] The dielectric function parameterization of BiFeO$_3$ is represented for this film by a Lorentz oscillator, three Tauc-Lorentz oscillator sharing a common Tauc gap, and a constant additive term to $\varepsilon_1$ represented by $\varepsilon_\infty$.[20,21] Figure 2 (a) shows an example of experimental transmittance spectra at T = 5 K for the 100 nm BiFeO$_3$ / DyScO$_3$ substrate stack and the fit using the parameterized dielectric function model described above. It displays an absorption onset at ~2.2eV, a small shoulder centered at $E_{TL}$~2.5eV, deriving from onsite $d$-to-$d$ excitations of the Fe$^{3+}$ ions and two larger features near 3.2 and 4.5eV that are assigned as charge transfer excitations.[22,23] There is good agreement between the experimental spectra and the fit. Similar fits were obtained for experimental spectra obtained over the full temperature range. Figure 2 (b) shows the position of the $E_{TL}$ ~2.5eV shoulder extracted from a parameterization of the absorption spectrum as a function of temperature. The raw data for the temperature-dependent absorption spectra can be found in reference 24. As observed, it shows singularities at $T_1^*$~380 K, $T_5^*$~580 K and $T_N$, consistent with the previously reported band gap temperature dependence.[24,25] and the current Raman results. Furthermore, from the results displayed in Fig. 2 (b), it can be seen that all the singularities observed in the magnetic Raman response (Figure 1(e)) occurs within the energy range of $E_{TL}$ ~ 2.43-2.34eV. In terms of $E_{TL}$, the first magnon anomaly at $T_1^*$~2.43eV is located ~0.11eV below the Raman pump energy ($E_{exc}$= 2.54eV), while the $T_5^*$~2.34eV is located ~0.2eV below the Raman pump. These values are close to the expected one-magnon energy ~0.1eV and the observed two-magnon energy ~0.19eV in Fig. 1. Further, the energy separation between $T_1^*$ and the others four transitions in terms of the corresponding shift in $E_{TL}$ are on the 20-50meV scale. We note that even when a detailed correspondence with specific phonon or magnon modes in BiFeO$_3$ is presently unwarranted, these energies are clearly on the lattice (phonon and magnon) energy scale.[16] Therefore, the above analysis suggests that the observed Raman enhancements arise from collective interactions involving (multi)-magnons, phonons and electronic states. Indeed, an additional signature for such phonon-magnon interaction at these resonances can be observed in the broad background peak that appears only at these magnon resonances (See Fig. 1(b)). Such broad features have also been seen in the Raman signal of multiferroic YMnO$_3$ and are attributed to two-magnon-phonon interactions, where phonons act as intermediate states.[26]

In summary, we have shown experimental evidence for spin-charge-lattice coupling in multiferroic BiFeO$_3$. Using a near resonant excitation wavelength (2.54 eV), six Raman enhancements of 2-magnon excitations are observed with temperature: one at $T_N$ arising from the antiferromagnetic phase transition, and 5 new ones below $T_N$, which are shown to be combined resonances involving electronic levels, phonons and magnon states. This study has broad implications for studying spin-charge-lattice coupling in multiferroics, especially where distinguishing polar and magnetic symmetries without neutron diffraction is a challenge. The work highlights the sensitivity of multimagnon spectroscopy to spin-charge coupling in multiferroics that has been minimally explored so far and appears to be broadly applicable to other multiferroics and magnetoelectrics.

We acknowledge funding from the National Science Foundation grant numbers DMR-0512165, DMR-0507146, DMR-0820404, DMR-0602986, DMR-0520513, and DMR-0520471 the MSD, BES U.S. Department of Energy under Contracts No. DE-AC02-05CH11231 and DE-FG02-01ER45885 and the DOE/BES under Contract No. DE-AC02-06CH11357.

**FIGURE CAPTIONS**

**Figure 1. (Color on line) (a)** Temperature-dependent Raman spectra of a 4.5 µm thick epitaxial BiFeO$_3$ film. The three Gaussian fits to RT spectra correspond to 2-phonon, 2-magnon, and 3-magnon replica, respectively, from low-to-high wavenumber shift. For clarity, the spectra at $T_1^*$ = 365 K, $T_2^*$ = 450 K, $T_3^*$ = 490 K, $T_4^*$ = 530 K, $T_5^*$ = 600 K, are normalized and vertically translated. **(b)** Evolution of the Raman spectra collected around $T_1^*$ **(c,d)** RT mid-infrared linear absorption measured on a BiFeO$_3$ bulk single crystal. The energy positions of two- and three-magnon absorption bands have been marked in the figure. **(e)** Integrated intensity ratio between the two-magnon and two-phonon Raman overtones as a function of temperature. $T_i^*$ ($i$=1-5) and $T_N$ denote transition temperatures.

**Figure 2. (Color on line) (a)** Absorption spectrum obtained for a 100nm BiFeO$_3$ film on DyScO$_3$(110). Solid lines are fits to the parameterized dielectric function at 5 K. **(b)** (Left axis) Energy position, $E_{TL}$ as a function of temperature. Also shown are resonance conditions at $T_i^*$ ($i$=1 to 5) and $T_N$ involving the Raman excitation wavelength $E_{exc}$, Tauc Lorentz energy, $E_{TL}$, one- and two-magnon and phonons $\Omega_1$, $\Omega_2$, and $\Omega_3$. (Right axis) Integrated intensity ratio between the two-magnon and two-phonon Raman overtones as a function of temperature.

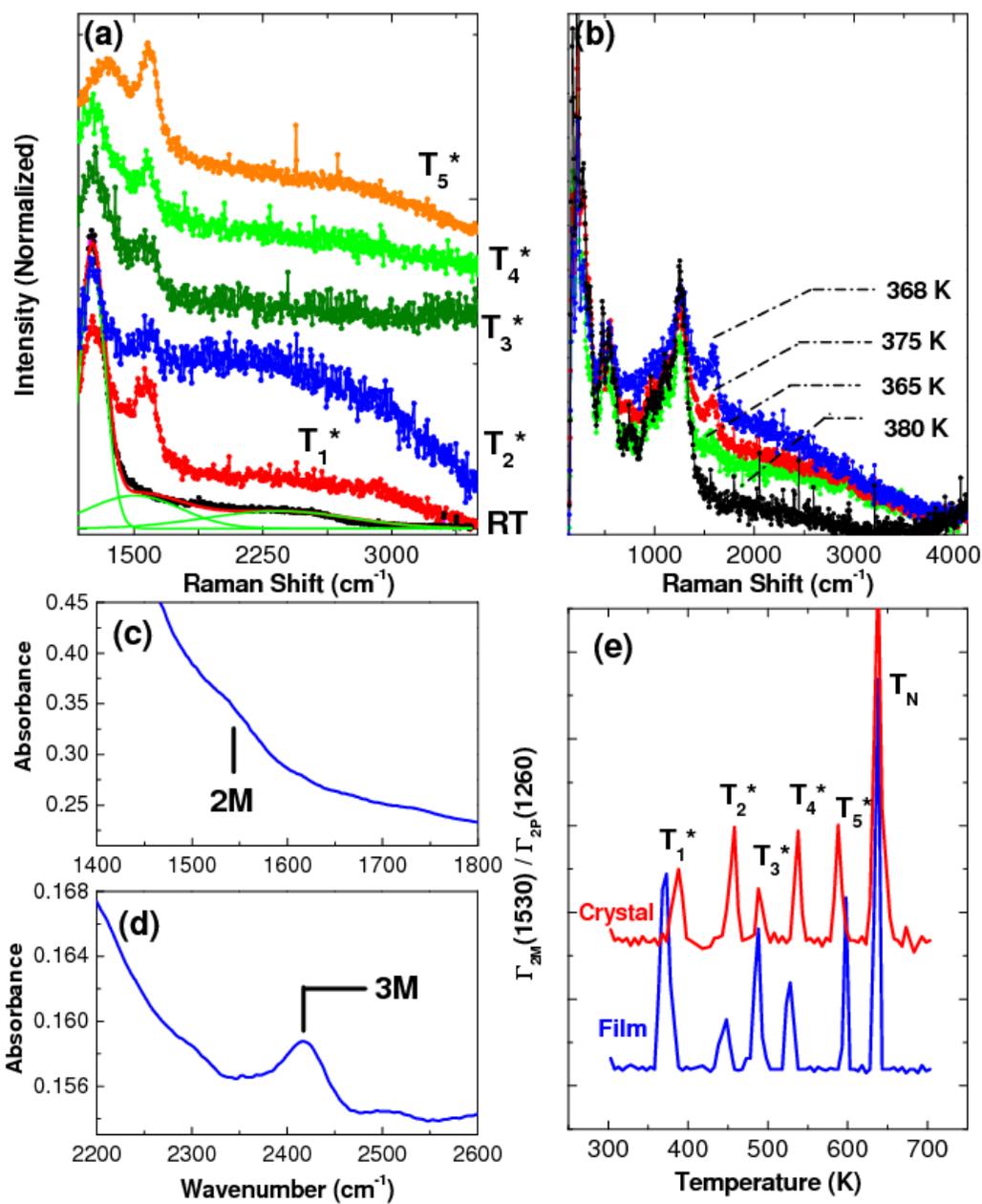

**Figure 1**

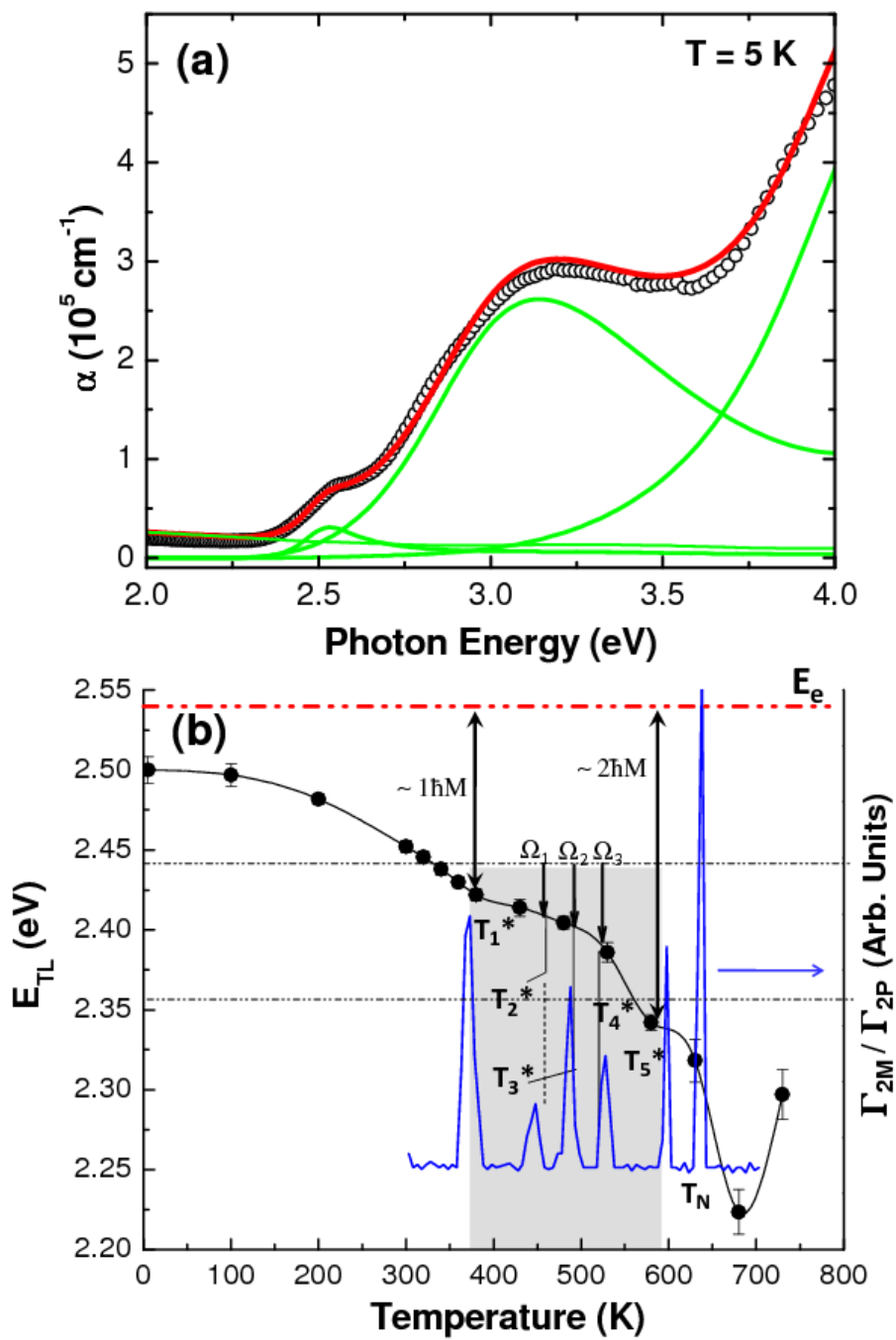

**Figure 2**